\RequirePackage{fix-cm}
\documentclass[3p]{elsarticle}
\usepackage{graphicx}
\usepackage{color}
\usepackage{caption}
\usepackage{subcaption}
\usepackage{minipage}
\usepackage{algorithm}
\usepackage{latexsym}
\usepackage{graphicx}
\usepackage{multicol}
\usepackage{amsmath,amssymb}
\usepackage{algorithm}
\usepackage{array}
\usepackage{eqparbox}
\usepackage{url}
\usepackage[bookmarks=true%
,bookmarksnumbered=true%
,hypertexnames=false%
,colorlinks=true%
,linkcolor=blue%
,citecolor=blue%
,urlcolor=blue%
]{hyperref}
\usepackage{algpseudocode}
\usepackage{float}
\usepackage{longtable,booktabs}
\newcommand{\delete}[1]{}

\newcolumntype{L}[1]{>{\raggedright\arraybackslash}p{#1}}

\journal{Information Processing and Management}
\begin{document}

\begin{frontmatter}


\title{A Deep Network Model for Paraphrase Detection in Short Text Messages} 




\author[affil1,affil2]{Basant Agarwal\corref{cor1}}
\cortext[cor1]{Corresponding author}
\ead{basant.agarwal@ntnu.no}

\author[affil1]{Heri Ramampiaro}
\ead{heri@ntnu.no}
\author[affil1]{Helge Langseth}
\ead{helgel@ntnu.no}
\author[affil1,affil3]{Massimiliano Ruocco}
\ead{massimiliano.ruocco@ntnu.no}

\address[affil1]{Dept. of Computer Science, Norwegian University of Science and Technology, Norway}
\address[affil2]{Swami Keshvanand Institute of Technology, India}
\address[affil3]{Telenor Research, Trondheim, Norway}


\date{Received: date / Accepted: date}


\begin{abstract}

This paper is concerned with paraphrase detection. The ability to detect similar sentences written in natural language is crucial for several applications, such as text mining, text summarization, plagiarism detection, authorship authentication and question answering. 
Given two sentences, the objective is to detect whether they are semantically identical. 
An important insight from this work is that existing paraphrase systems perform well when applied on clean texts, but they do not necessarily deliver good performance against noisy texts. Challenges with paraphrase detection on user generated short texts, such as Twitter, include language irregularity and noise. To cope with these challenges, we propose a novel deep neural network-based approach that relies on coarse-grained sentence modeling using a convolutional neural network and a long short-term memory model, combined with a specific fine-grained word-level similarity matching model. Our experimental results show that the proposed approach outperforms existing state-of-the-art approaches on user-generated noisy social media data, such as Twitter texts, 
and achieves highly competitive performance on a cleaner corpus. 
 
\end{abstract}


\end{frontmatter}

\section{Introduction}

Twitter has for some time been a popular means for expressing opinions about a variety of subjects. Paraphrase detection in user-generated noisy texts, such as Twitter texts\footnote{From now on referred to as Tweets.}, is an important task for various Natural Language Processing (NLP), information retrieval and text mining tasks, including query ranking, plagiarism detection, question answering, and document summarization. Recently, the paraphrase detection task has gained significant interest in applied NLP because of the need to deal with the pervasive problem of linguistic variation.



Paraphrase detection is an NLP classification problem. Given a pair of sentences, the system determines the semantic similarity between the two sentences. If the two sentences convey the same meaning it is labelled as \emph{paraphrase}, otherwise \emph{non-paraphrase}. Most of the existing paraphrase systems have performed quite well on clean text corpora, such as the Microsoft Paraphrase Corpus (MSRP)~\cite{Dolan:2004:UCL:1220355.1220406}. However, detecting paraphrases in user-generated noisy tweets is more challenging due to issues like misspelling, acronyms, style and structure~\cite{Xu_2014}.
Although little attention has been given to paraphrase detection in noisy short-texts, some initial work has been reported on the benchmark \emph{SemEval 2015} Twitter dataset~\cite{Xu_2014,Xu_2015,Dey2016}. 
Unfortunately, the best performing approaches on one dataset perform poorly when evaluated against another. 
As we discuss later in this paper, the state-of-the-art approach for the SemEval dataset proposed by \citet{Dey2016} gives quite poor F1-score when evaluated on the MSRP dataset. Similarly, \citet{DBLP:conf/emnlp/JiE13} is the best performing approach on  the MSRP dataset, but does not perform well on the SemEval dataset. 
In conclusion, existing approaches are not very generic; but instead, they are highly dependant on the data used for training. 

Focusing on the problem discussed above, the main goal of this work is to develop a robust paraphrase detection model based on deep learning techniques that is able to successfully detect paraphrasing in both noisy and clean texts. 
More specifically, we propose a hybrid deep neural architecture composed by a \textit{convolutional neural network} (CNN) and a \textit{long short-term memory} (LSTM) model, further enhanced by a novel word-pair similarity module. The proposed paraphrase detection model is composed of two main components, i.e., {\em pair-wise word similarity matching} and {\em sentence modelling}. The pair-wise similarity matching model is used to extract fine-grained similarity information between pairs of sentences. 
We use a CNN to learn the patterns in the semantic correspondence between each pair of words in the two sentences that are intuitively useful for paraphrase identification. 
The idea to apply convolutions over a pair-wise word to word similarity matrix to extract the important word-word similarity pairs is motivated by how convolutions over text can extract the most important parts of a sentence. 
In sentence modelling architecture, we extract the local region information in form of important n-grams from the text using the CNN, and the long-term dependency information using the LSTM. By using this architecture, we are able to develop an informative semantic representation of each sentence. In this paper, we show how the proposed model can be enhanced by employing an extra set of statistical features extracted from the input text. 
To demonstrate its robustness, we evaluated the proposed approach and compare it with the state-of-the-art models, using two different datasets,  covering both noisy user-generated texts - i.e., the \emph{SemEval 2015} Twitter benchmark dataset, and clean texts - i.e., the Microsoft Paraphrase Corpus (MSRP).

 
In summary, the main contributions of this paper are:
\begin{enumerate}
  \item We propose a novel deep neural network architecture leveraging coarse-grained sentence-level features and fine-grained word-level features for detecting paraphrases on noisy short text from Twitter. The  model combines sentence-level and word-level semantic similarity information such that it can capture semantic information at each level. When the text is grammatically irregular or very short, the word-level similarity model can provide useful information, while the semantic representation of the sentence provide useful information otherwise. In this way both model-components compliment each other and provide efficient overall performance.
  \item We show how the proposed pair-wise similarity model can used to extract word-level semantic information, and demonstrate its usefulness in the paraphrase detection task.
  \item We propose a method combining statistical textual features and features learned from the deep architecture.
  \item We present an extensive comparative study for the paraphrase detection problem. 
\end{enumerate}

The rest of the paper is organized as follows: 
We formally define the problem  in Section~\ref{problem},
then discuss related word concerning paraphrase detection in Section~\ref{related}. 
In Section~\ref{proposed}, we motivate our work and present our proposed solution in detail. 
Thereafter, we describe the experimental setup
in Section~\ref{eval}, and evaluate the approach and discuss the results 
in Section~\ref{results}. 
Finally, in Section~\ref{conclusions}, we conclude the paper and outline plans for future research.

\section{Problem statement}
\label{problem}

Let $S_1$ and $S_2$ be two sentences, such that $S_1 \neq S_2$. 
$S_1$ and $S_2$ are said to be paraphrased if they convey the same meaning and are semantically equivalent. 
Now, assume that we have a collection of $N$ annotated sentence pairs ($S_1^i$, $S_2^i$), having annotations ${k_i}$, for $i$ = $1, 2, \ldots N$. 
For a given $i$, ${k_i}$ indicates whether the $i$-th sentence pair is {\em paraphrased} or {\em non-paraphrased}. 
The problem addressed in this paper is to develop a model, which can reliably annotate a previously unseen sentence pair as paraphrased or non-paraphrased. 

There are several methods that have been proposed, and work well for clean texts, but most of them have failed to provide satisfactory results when applied on noisy texts like Tweets. 
On the other hand, some approaches have recently been developed for paraphrase detection on noisy texts, e.g., the work by \citet{Xu_2014} and \citet{Dey2016}, but as shown later, these approaches do not work well on clean texts. In conclusion, there is still a strong need for a robust and reliable method, which can perform well for both clean texts and user-generated noisy short texts. Addressing this need is the main objective of the work presented in this paper.

\section{Related work}
\label{related}

The use of deep neural network for natural language processing has increased considerably over the recent years. 
Most of the previous work on sentence modelling have focused on feature  like n-gram overlap features~\cite{Madnani:2012:RMT}, syntax features~\cite{Rus2008,Das:2009:PIP}, and machine translation based features~\cite{Madnani:2012:RMT}. 
Recently, deep learning-based methods have shifted researchers' attention towards semantically distributed representations. 
A variety of  deep neural network-based architectures have been proposed for sentence similarity, a strategy we also focus on in this paper.

Substantial work has been carried out on paraphrase detection from the clean-text Microsoft Paraphrase corpus. 
\citet{Das:2009:PIP} present a probabilistic model for paraphrase detection based on syntactic similarity, semantics, and hidden loose alignment between syntactic trees of the two given sentences. 
\citet{Heilman2010} propose a tree edit model for paraphrase identification  based on syntactic relations among words. They develop a logistic regression model that uses 33 syntactic features of edit sequences to classify a sentence pair. 
\citet{Socher:2011} present an approach based on recursive autoencoders for paraphrase detection. Their approach learns feature vectors for phrases in syntactic trees and employs a dynamic pooling layer mechanism, which converts a variable sized matrix into a fixed-sized representation. Parsing is a powerful tool for identifying the important syntactic structure in the text, but relying on the parsing makes the approach less flexible. Our approach does not use such resources to develop the model. \citet{OLIVA2011} propose SyMSS based on the syntactic structure of the sentences. They represent the sentences as a syntactic dependence tree, use WordNet to extract meaning of individual words, and further use syntactic connections among them to assess information similarity. \citet{DBLP:conf/emnlp/JiE13} use several hand-crafted features with latent representation from matrix factorization as features to train a support vector machine. 
The ARC model proposed by \citet{NIPS2014_5550} is a convolutional Siamese architecture, in which two shared-weight convolutional sentence models are trained. 
\citet{El-Alfy:2015} propose a  model considering a set of weak textual similarity metrics. They boost the performance of individual metrics using abductive learning. Further, they aim to select an optimal subset of similarity measures and construct a composite score that is used for classification.
\citet{Wang2016SentenceSL} decompose the sentence similarity matrix into a similar component matrix and a dissimilar component matrix, and train a two-channel convolutional neural network to compose these components into feature vectors.  \citet{FERREIRA201859} propose a supervised machine learning learning approach.
They extract various features based on lexical, syntactic and semantic similarity measures, and use various machine learning algorithms such as Bayesian Network, RBF Network, C4.5 decision tree , and support vector machines.
 
Some contributions have also been reported for detecting paraphrases on noisy short-text like Tweets. \citet{Xu_2014} propose a latent variable model that jointly infer the correspondence between words and sentences. \citet{Eyecioglu} use a support vector machine with simple lexical word overlap and character n-grams features for paraphrase detection. \citet{Zhao2015} use various machine learning classifiers, and employ a variety of features like string-based, corpus-based, syntactic features,  and word distributional representations. 
\citet{Zarrella2015} present an ensemble approach based on various features such as  mixtures of string matching metrics, distance measurements, tweet-specific distributed word representations,  and recurrent neural networks for modeling similarity. 
\citet{Karan2015} present a supervised approach that combines semantic overlap and word alignment features. 
\citet{Vo} experiment with various sets of features with different classifiers and show that the combination of word/n-gram, word alignment by METEOR (Metric for Evaluation of Translation with Explicit ORdering), BLEU (Bilingual Evaluation Understudy) and EditDistance is the best feature set for Twitter paraphrase detection; VotedPerceptron proved to be the best machine learning algorithm. 
\citet{Dey2016} use a set of lexical, syntactic, semantic and pragmatic features. 

In this paper, we focus on using deep learning algorithms to develop a robust and reliable paraphrase detection system, which can work well on both clean-text and noisy short text such as tweets. To the best of our knowledge, this is the first work to fully explore this area, while also including a comprehensive comparative study of exiting approaches. 

Table~\ref{tab:comparison-related-work} summarizes the approaches discussed in section. 

 
 \small
\begin{longtable}
{ p{0.8cm} L{4.5cm} L{3.5 cm} p{2.8cm} p{1.1cm}} 
\caption{Comparison among related approaches. }
\label{tab:comparison-related-work} \\
\toprule
{\bf Work}  & {\bf Description} & {\bf Resources used}  & {\bf Classification}  & {\bf dataset} \\ 
\midrule
\endfirsthead
\caption{Comparison among related works (cont.).}
\label{table10}        \\
\toprule
{\bf Work}  & {\bf Description} & {\bf Resources used}  & {\bf Classification}  & {\bf dataset} \\ 
\midrule
\endhead
    \multicolumn{5}{r}{\footnotesize\itshape Continue on the next page.}
\endfoot
    \bottomrule
\endlastfoot    
    
\cite{Eyecioglu}  &ASOBEK: Word overlap and character n-grams features  & POS tagger & Support vector machine (SVM)  &  Twitter, MSRP \\
\midrule
\cite{Zarrella2015} & MITRE: mixtures of  string  matching  metrics & --  &  L1-regularized logistic regression  &  Twitter dataset  \\
\midrule
 \cite{Zhao2015} & ECNU: Various string based, corpus based, syntactic, and distributed word representation based features & POS tagger, WordNet, various pre-trained word embeddings & SVM, Random Forest (RF), Gradient Boosting (GB)  & Twitter dataset   \\
\midrule
\cite{Vo}   &  Various features such as Machine translation, EDIT distance, sentiment features  & POS Tagger & Decision Stump, OneR, J48, Baysian Logistic Regression, VotedPerceptron, MLP & Twitter dataset \\
\midrule
\cite{Karan2015} &  Semantic Overlap Features and  Word Alignment Features &POS tagger &SVM  & Twitter dataset\\
\midrule
\cite{Xu_2014}  &  Multi-instance Learning Paraphrase Model (MULTIP) & POS tagger &  Similarity score   &   Twitter dataset  \\
\midrule
\cite{Dey2016}  & A set of lexical, syntactic, semantic and pragmatic features & WordNet, POS Tagger, NE Tags &   SVM &  Twitter, MSRP    \\
\midrule
 \cite{Mihalcea2006} & Combination of several word similarity measures & POS tagger & Similarity score threshold & MSRP  \\
\midrule
 \cite{Guo:2012}  & Weighted Textual Matrix Factorization (WTMF) with handling missing words & WordNet &  Matrix factorization & MSRP \\
 
\midrule
\cite{Das:2009:PIP} & Probabilistic model with syntactic and n-gram overlap features & WordNet, Dependency parser &  Logistic regression, SVM & MSRP\\
\midrule
\cite{Heilman2010} & Syntactic features of edit sequences & POS Tagger, Parser, WordNet & Logistic regression & MSRP \\
\midrule
\cite{OLIVA2011} & Similarity features based in syntactic dependency tree &  WordNet, dependency parser &  Similarity score threshold & MSRP\\
\midrule
\cite{Socher:2011} & Representation of feature vectors for phrases in syntactic trees  &  Dependency Parser & Recursive autoencoder with dynamic pooling & MSRP \\
\midrule
\cite{DBLP:conf/emnlp/JiE13}  &  Matrix factorization with supervised reweighting & -- & SVM with a linear kernel & MSRP \\
\midrule
 
\cite{NIPS2014_5550}  & Hierarchical structures of sentences with their layer-by-layer composition  & pre-trained word embeddings & Convolutional Neural Network  & MSRP  \\
\midrule
 
\cite{Madnani:2012:RMT} & Combination of eight machine translation metrics  & WordNet & SVM  & MSRP  \\
\midrule
\cite{El-Alfy:2015} & Boosting through textual similarity metrics & -- & SVM  & MSRP\\
\midrule
\cite{Wang2016SentenceSL} & Sentence Similarity Learning by Lexical Decomposition and Composition  & pre-trained word Embeddings& CNN & MSRP\\
\midrule

\cite{FERREIRA201859}   & Represent pair of sentence as combination of similarity measures & Dependency Parser & SVM, RBF Network, Bayesian Network &  MSRP \\
\midrule
This work  &     Hybrid of deep learning and statistical features & POS Tagger \& pre-trained word Embeddings & Multi-layer neural network & MSRP, Twitter dataset\\

\end{longtable}
\normalsize


\section{DeepParaphrase Architecture}
\label{proposed}



\begin{figure*}[htbp]
\hspace*{-0.1in}
\centering
\includegraphics[scale=0.55]{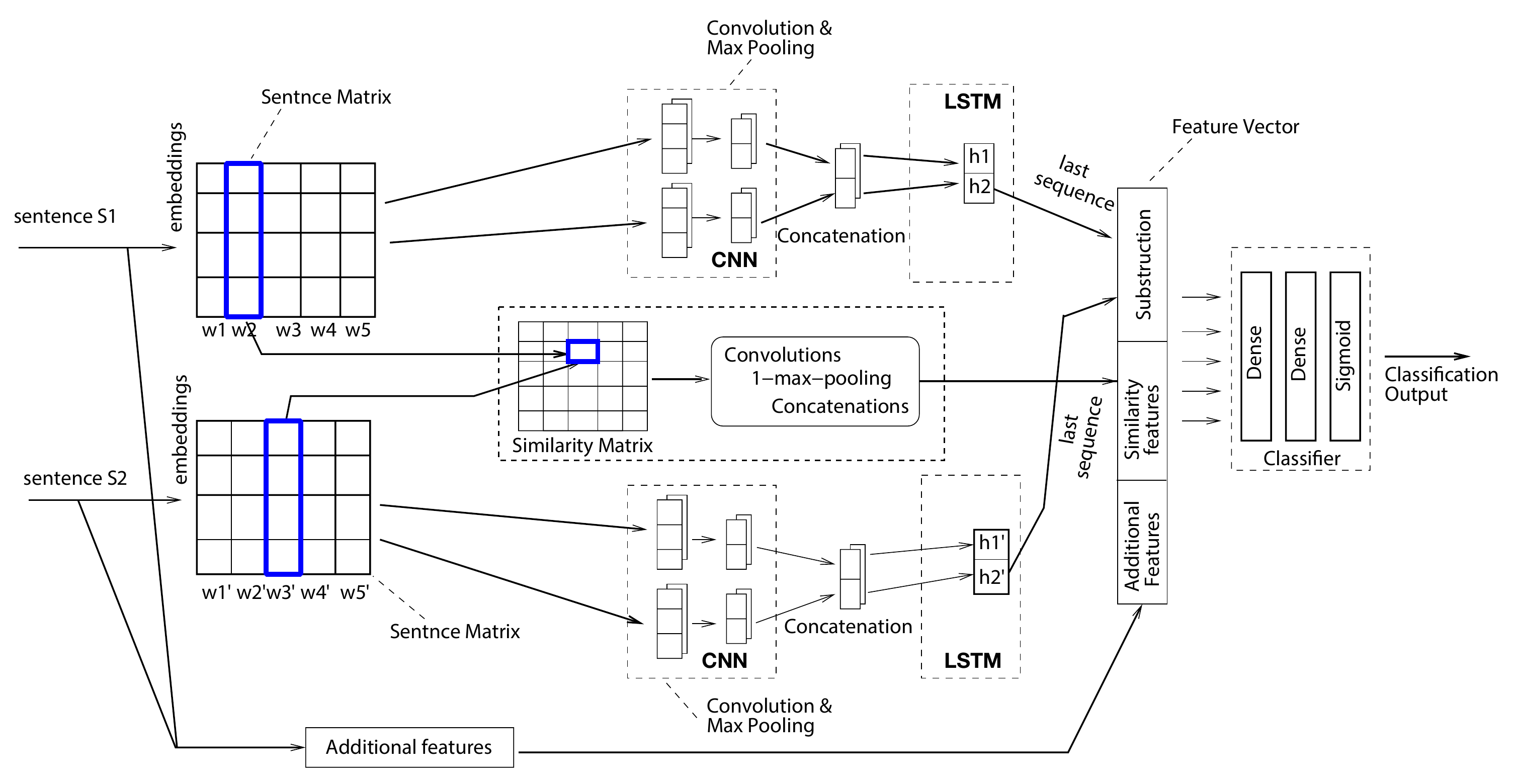}
\caption{The proposed DeepParaphrase architecture}
\label{systemfig}
\end{figure*}

We propose a deep learning-based approach for detecting paraphrase sentences for tweets. 
We first convert each sentence in a pair into a semantic representative vector, using a CNN and an LSTM. 
Then, a semantic pair-level vector is computed by taking the element-wise difference of each vector in the sentence representations. 
The resulting difference is the discriminating representative vector of the pair of sentences, which is used as feature vector for learning the similarity between the two sentences. 
In addition to this coarse-level semantic information, we extract more fine-grained important information using a similarity matrix which contains word-to-word similarity quantification.
Further convolutions are applied over the pair-wise similarity matrix to learn the similarity patterns between the words in the pair of sentences. 
The aim of the convolution function is to extract more fine-grained similarity features. Finally a third set of features are extracted  using statistical analysis of the text, and concatenated with the rest of the learned features. 
A fully connected neural network 
is used to 
produce the classification from this concatenated feature vector. 
The first layers are activated by the \emph{ReLU}~\cite{ICML-2010-NairH} function, while we use the sigmoid link-function to transfer the latent representation into a two-class decision rule. We train the model to optimize  binary cross-entropy.
The proposed architecture is depicted in Figure~\ref{systemfig}.

At a high level of abstraction the proposed model therefore consists of two main components, that will 
be discussed next.

\delete{
\begin{enumerate}
    \item \emph{\textbf{Sentence modeling with CNN and LSTM}}:  In this component, we represent every sentence using our joint CNN and LSTM architecture. 
    The CNN is able to learn the local features from words to phrases from the text, while the LSTM learns the long-term dependencies of the text. 
    More specifically, we firstly take the word embedding as input to our CNN model, in which various types of convolutions and pooling techniques are applied to capture the maximum information from the text. Next, the encoded features are used as input to the LSTM network. Finally, the long term dependencies learned by the LSTM becomes the semantic sentence representation.
    
    \item \emph{\textbf{Pair-wise word similarity matching}}: A pair-wise similarity matrix is construed by computing the similarity of each word to another word in both sentences. Convolutions are applied onto this similarity matrix to analyze patterns in the pair-wise word to word similarities. 
\end{enumerate}

The aforementioned components are further detailed in the following sections.
}

\subsection{Sentence modelling with CNN and LSTM}

In this component, we represent every sentence using our joint CNN and LSTM architecture. 
The CNN is able to learn the local features from words to phrases from the text, while the LSTM learns the long-term dependencies of the text. 
More specifically, we firstly take the word embedding as input to our CNN model, in which various types of convolutions and pooling techniques are applied to capture the maximum information from the text. Next, the encoded features are used as input to the LSTM network. Finally, the long term dependencies learned by the LSTM becomes the semantic sentence representation.

The architecture of the proposed model for mapping the sentences into a feature vector is shown in Figure~\ref{systemfig}. 
The main goal of this step is to learn good intermediate semantic representations of the sentences, which are further used for the semantic similarity task. 
%
%
The input to the sentence model is a pair of sentences $S_1$ and $S_2$, which we transform into matrices of their words' embeddings. Here each  word is represented by a vector $w \in \mathbb{R}^{d}$, where  $d$ is the  size of the word embedding. 
We used pre-trained word embeddings (see Section~\ref{hyperp} for details).
The sentence embedding matrices are then fed into the CNN. The result captures the local region information, and is used as input to the LSTM. 
The aim of the convolutional layer is therefore to extract patterns, i.e., important word sequences from the input sentences. 
The motivation for using convolutions comes from the fact that convolutional filters can learn n-gram discriminating features, which is useful for  sentence similarity analysis. 

The features generated by the convolutional layer have the form of n-grams, and are fed into the LSTM. This model component is able to process sequential input with the aim to learn the long-term dependencies in the sentences.
Eventually, the \textit{last} latent layer of the LSTM is taken as the semantic representation of the sentence, and the difference between element-wise difference between these representations is used as a semantic discrepancy measure at the level of the sentence pair.

\subsection{Pair-wise word similarity matching}
\begin{figure}[htbp]
 
\centering
\includegraphics[scale=0.45 ]{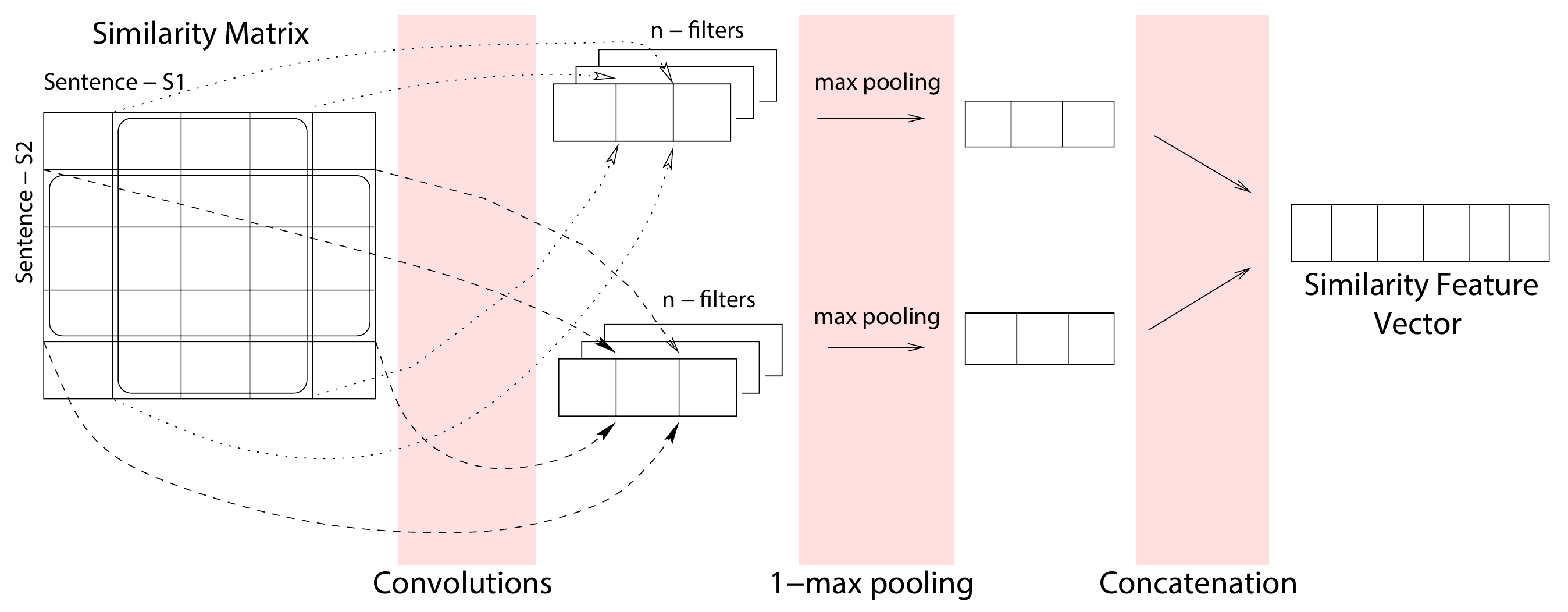}
\caption{Pair-wise word similarity matching model}
\label{system:fig2}
\end{figure}

A pair-wise similarity matrix is construed by computing the similarity of each word in $S_1$ to another word in $S_2$. Convolutions are applied onto this similarity matrix to analyze patterns in the pair-wise word to word similarities. Figure~\ref{system:fig2} illustrates this process.

It is intuitive that given two sentences, semantic correspondence between words provide important semantic information for detecting similar sentences, and the pair-wise word similarity matching model learns the word-level similarity patterns between the two sentences. 
Because important n-grams are extracted by applying convolutional neural network over text, we obtain the important word-word similarity pairs from the similarity matrix. This similarity matrix is further used as features for the classification of the paraphrase detection problem. The goal of the pair-wise word similarity matching model is to compare the semantic embedding of each word in one sentence against all the semantic embeddings of the words from the other sentences.
This means that we compute the dot product as a similarity measure between all the word embeddings of the two sentences.
Finally, we match the two sentences and generate a similarity matrix  $\mathbb{S}$ of size $m \times n$, where $m$ and $n$ denote the lengths of sentence $S_1$ and $S_2$, respectively. 
Next, we apply the CNN onto the similarity matrix to learn the patterns in the semantic  correspondence between the two sentences.
We convolve over $\mathbb{S}$ in two directions; both from left to right and from top to bottom. This gives two separate results, $F_1$ and $F_2$.
After the convolution layer, global max-pooling is applied to obtain the most informative feature vectors from $F_1$ and $F_2$,
and finally these are concatenated to produce the output from this module.


\delete{
\begin{algorithm*}
    \caption{Pair-wise Word Similarity Matching }\label{pairwise}
      \textbf{Input}: \emph{Sentence matrices $S1 \in R^{d \times m}$ and $S2 \in R^{d \times n}$ for each sentence.} \\
     \textbf{Output}: \emph{ Fine-grained word-word similarity feature vector Q}
    \begin{algorithmic}[1]

   \For{\emph{ 1 to m }}
        \For{\emph{1 to n}}
            \State \emph{ Sim[i][j] $\gets$ S1[i] $\cdot$ S2[j]. }  \emph{//Construct similarity matrix $(Sim) \in R^{m \times n}$}
            
        \EndFor
    \EndFor
    
    \For{\emph{each filter width $f_i$}}
            \State \emph{Feature map matrix $C1 \in R^{k \times (m-p+1)}$ } $ \gets$ \emph{Convolve weight matrix $M \in R^{d \times p}$ over Sim.}
            \State \emph{Apply Global max-pooling} $F1$ $\gets$ $R^{k}$ 
    \EndFor
    \State \emph{TransSim $\gets$ Transpose Sim} 
    \For{\emph{each filter width $f_i$}}
                \State \emph{Feature map matrix $C2 \in R^{k \times (m-p+1)}$ } $ \gets$ \emph{Convolve weight matrix $M \in R^{d \times p}$ over TransSim.}
            \State \emph{Apply Global max-pooling} $F2$ $\gets$ $R^{k}$ 
    \EndFor
    \State \textit{Concatenation: Similarity feature vector $Q \in R^{2k}$} $\gets$  $F_1$ $\oplus$  $F_2$

\end{algorithmic}
\end{algorithm*}
}

\subsection{Statistical features}
\label{statsF}
We extracted a third set of features
to enhance the discriminating representation of the sentences. These features consist of the following: 

\begin{enumerate}
    \item \emph{TF-IDF}-based similarity between sentences $S_1$ and $S_2$.
    \item Cosine similarity between the vectors of sentences $S_1$ and $S_2$.
    \item The average Wordnet-based similarity between the verbs\footnote{\label{foot:nltk} In our implementation we use \texttt{nltk}'s part of speech tagger to extracts the verbs, nouns and adjectives from each sentence.} in sentence $S_1$ and those in $S_2$.
 
    \item The average Wordnet-based similarity between the nouns\textsuperscript{\ref{foot:nltk}} in sentence $S_1$ and $S_2$.
    \item The average Wordnet-based similarity between the adjectives\textsuperscript{\ref{foot:nltk}} in sentence $S_1$ and $S_2$.
    \item \label{item:similarity-feature}
    The cosine similarity between the semantic representation of each sentence pair. 
    \item \label{item:overlap-feature} 
    Six n-gram overlap features computed by the number of unigrams, bigrams, and trigrams that are common to the given sentence pair, divided by the total n-grams in $S_1$ and $S_2$ respectively.  
    
\end{enumerate}

We use all these additional features for the experiments performed on the Microsoft Paraphrase corpus, while only the two latter features were used for the experiments on Twitter corpus.



\section{Experimental Setup}
\label{eval}

Before evaluating our proposed method for  paraphrase identification, and compare it against the state-of-the-art approaches, we first describe how our experiments have been set up, including the datasets, performance measures and the hyperparameter settings that we have used.

\subsection{datasets} 
\label{sec:eval-datasets}
We consider two widely-used benchmark datasets, which we briefly describe in the following: 

\begin{enumerate}
  \item  \textbf{Twitter Paraphrase SemEval 2015 dataset:} The dataset provided by SemEval 2015 \cite{Xu_2015} has been used by all the recent works for paraphrase detection in Tweets. 
  It consists of noisy and short text, containing 3996 paraphrase and 7534 non-paraphrase pairs in the training dataset, 1470 paraphrase and 2672 non-paraphrase sentence pairs in development set, and
  838 tweets in the test set. We have ignored the ``debatable'' entries, that were marked in~\cite{Xu_2015}. The statistics of the dataset are shown in Table~\ref{tab:dataset-stats}.

\begin{table*}[htbp]
\caption{Statistics of the SemEval-2015 Twitter paraphrase corpus}
\label{tab:dataset-stats}
\begin{center}
\begin{tabular}{llllll}
\toprule \bf   & \bf Unique sent.  & \bf Sent. pair & \bf Paraphrase & \bf Non-paraphrase & \bf Debatable \\ \midrule
Train & 13231 & 13063 & 3996 & 7534 & 1533\\
Dev & 4772 & 4727 & 1470 & 2672 & 585 \\
Test & 1295 & 972 & 175 & 663 & 134 \\
 
\bottomrule
\end{tabular}
\end{center}

\end{table*}

 \item \textbf{Microsoft Paraphrase dataset:} We also investigate the empirical performance of the proposed model on a clean text corpus. More specifically, we use the Microsoft Paraphrase dataset~\cite{Dolan:2004:UCL:1220355.1220406}, which is considered the evaluation standard for paraphrase detection algorithms. This dataset comprises candidate paraphrase sentence pairs, obtained from Web news sources. 
 In this corpus, the length of each sentence varied from 7 to 35 words with an average 21 words in a sentence, and $67\%$ of the sentence pairs are marked as paraphrased. 
 Furthermore, the data is split into training and test sets, containing 4076 and 1725 samples respectively. This same train/test partitioning has been applied on all the approaches evaluated in this paper.

\end{enumerate}
 
Despite being the most widely-used datasets for evaluating paraphrase detection models, their sizes are too small to reliably train a deep learning architecture. We have therefore applied a simple augmentation scheme to double the number of sentence pairs in the corpus: For every pair of sentences $(S_1, S_2)$ we simply exchange the order of sentences to obtain the new pair $(S_2, S_1)$, and add this new pair to the corpus.

\subsection{Performance measures}
\label{sec:performance-measures}

We adopted the standard performance measure that are widely used in the literature for paraphrase detection. These measures are \emph{precision}, \emph{recall}, \emph{F1-score} and \emph{accuracy}.  Precision is defined as number of  correctly classified paraphrase pairs out of total paraphrase sentence pairs extracted, hence  computed as 
\begin{equation}
\nonumber
\text{Precision} = \frac{TP}{TP+FP}.
\end{equation}
\noindent
Here, TP refers to True Positives, i.e., number of paraphrase pairs classified as paraphrase, while FP refers for False Positives, i.e., number of non-paraphrase pairs determined as paraphrase. 

Recall is the ratio between predicted sentence pairs that are actual paraphrases, and total true paraphrase pairs:
\begin{equation}
\nonumber
\text{Recall} = \frac{TP}{TP+FN}.
\end{equation}
\noindent
Here, FN is False Negatives, i.e., number of paraphrase pairs classified as non-paraphrase pairs); TN means True Negatives, i.e., number of non-paraphrases determined as non-paraphrases. 
The F1-score combines the precision and recall: 

\begin{equation}
\nonumber
 \text{F1-score} = \frac{2\cdot\text{Precision}\cdot\text{Recall}}{\text{Precision}+\text{Recall}}
\end{equation}
Finally, accuracy is the fraction of the paraphrase sentence pairs that are classified correctly:
\begin{equation}
\nonumber
\text{Accuracy} = \frac{TP+TN}{TP+TN+FN+FP}
\end{equation}

\subsection{Hyperparameter Setting}
 \label{hyperp}
 
Hyperparameters were chose by rough investigations into the training data to choose optimization algorithm, learning rates, regularization, and size of training dataset. The optimal settings for these hyperparameters vary between datasets, hence we choose separately for the Twitter and MSRP datasets.

\subsubsection{Hyperparameter settings on the Twitter dataset}

\delete{
{\Huge\bf This part is not correct. Delete ?}
We define all the hyperparameters to as the set $\Theta$, denote the set of training data as $\chi$, and the set of labels as  $\gamma$. 
For each $x \in \chi$, the model computes a score $s$ for each $y \in \gamma$. To transform the scores $s$ into a conditional probability distribution in the output layer, we use the softmax
\begin{equation}
p(y|x,\Theta) = \frac{\exp(s)}{\sum_{j}\exp(s_j)}
\end{equation}


}

\begin{figure*}[tb]
    \centering%
    \begin{subfigure}[t]{0.45\textwidth}%
        \includegraphics[width=\linewidth]{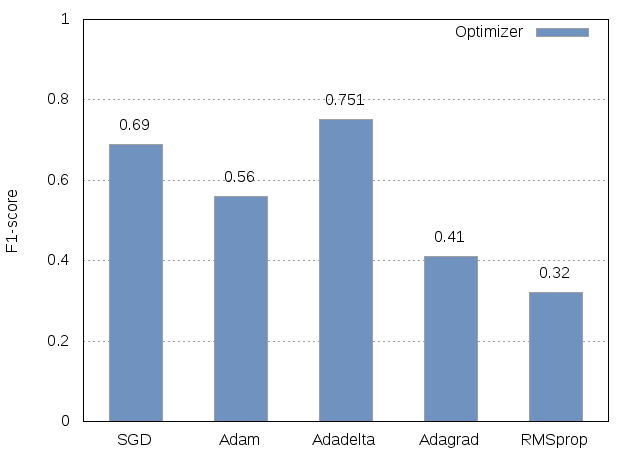}%
        \caption{Performance of optimization algorithms}\label{optimizer_Twitter}%
    \end{subfigure}%
    ~
    \begin{subfigure}[t]{0.45\textwidth}%
        \includegraphics[width=\linewidth]{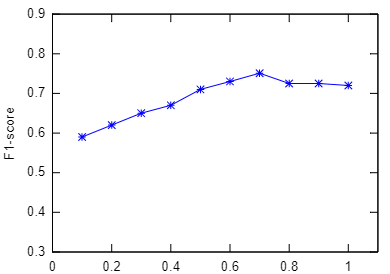}%
        \caption{Performance vs.\ learning rate}\label{fig:performance-diff-adadelta}%
    \end{subfigure}%
    \bigskip
    
    \begin{subfigure}[t]{0.45\textwidth}%
      \includegraphics[width=\linewidth]{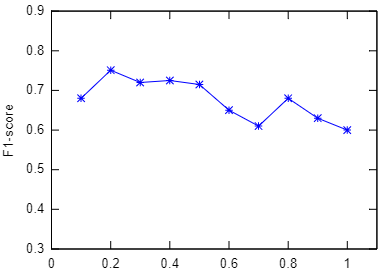}%
      \caption{Performance vs.\  dropout rate}\label{dropout_Twitter}%
    \end{subfigure}%
    ~
    \begin{subfigure}[t]{0.45\textwidth}%
        \includegraphics[width=\linewidth]{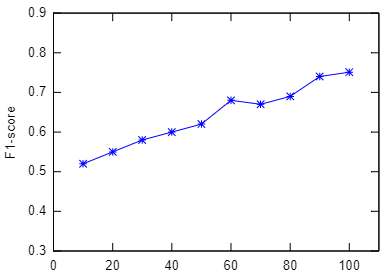}%
        \caption{Learning curve}\label{Fraction_Twitter}%
    \end{subfigure}%
    \caption{Evaluation of different hyperparameters for the SemEval Twitter dataset}
\end{figure*}

We empirically experiment with various optimizers, see Figure~\ref{optimizer_Twitter}, and chose \emph{Adadelta} to optimize the learning process. 
%
%
%
%
We further tune the learning rate for this optimizer, see Figure~\ref{fig:performance-diff-adadelta}, with  learning rate  0.70 appearing to be optimal. 
Dropout is used for regularization of the proposed model. This prevents feature co-adaptation by randomly setting a portion of the hidden units to zero during training. 
We applied dropout to every layer, and set the dropout ratio to 0.2, cf.\ Figure~\ref{dropout_Twitter}.
Finally, we investigate the sensitivity of the approach wrt.\ the amount of training data supplied. Figure~\ref{Fraction_Twitter} shows the learning curve, i.e., the learning quality as a function of the amount of the training data used. We clearly see an increasing trend in the learning curve, which indicates that more training data may further improve the performance of the proposed model.
%

%
  





In the absence of a large supervised training set, it is common to initialize word embeddings with pretraining values that have been obtained from an unsupervised neural language model~\cite{Collobert:2011:NLP}. We follow this strategy, and used the popular \texttt{glove} embeddings\footnote{The embeddings are available at \url{{https://nlp.stanford.edu/projects/glove/}}.}
during our experiments on Twitter dataset. We chose the embeddings  pretrained on 2 billion tweets, and use the 200-dimensional version.

\subsubsection{Hyperparameter settings for MSRP dataset}
\label{sec:hyperparameter-setting-MSRP}

\begin{figure*}[tb]
    \centering
    \begin{subfigure}[t]{0.45\textwidth}%
        \includegraphics[width=\linewidth]{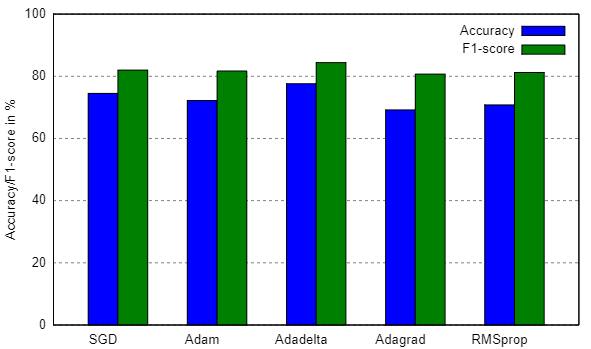}%
        \caption{Performance of optimization algorithms}\label{optimizer_msr}%
    \end{subfigure}%
    ~
    \begin{subfigure}[t]{0.45\textwidth}%
        \includegraphics[width=\linewidth]{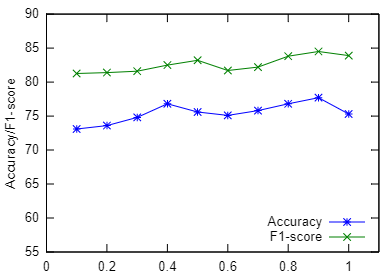}%
        \caption{Performance vs.\ learning rate}\label{fig:learning_msr}%
    \end{subfigure}%
    \bigskip
    
    \begin{subfigure}[t]{0.45\textwidth}%
      \includegraphics[width=\linewidth]{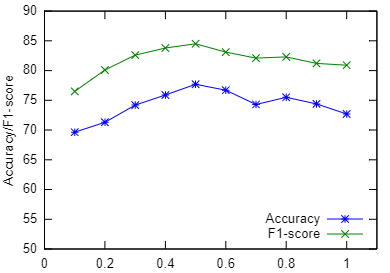}%
      \caption{Performance vs.\  dropout rate}\label{dropout_msr}%
    \end{subfigure}%
    ~ 
    \begin{subfigure}[t]{0.45\textwidth}%
        \includegraphics[width=\linewidth]{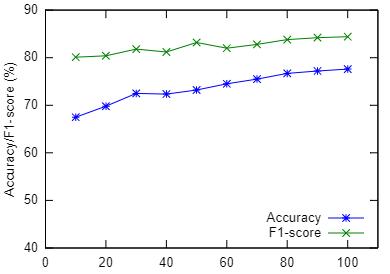}%
        \caption{Learning curve}\label{Fraction_msr}%
    \end{subfigure}%
    \caption{Evaluation of different hyperparameters for the MSRP dataset}\label{fig:hyper:msr}
\end{figure*}

The parameter selection process for the MSRP dataset is similar to what was discussed for the Twitter data above, see Figure~\ref{fig:hyper:msr} for results. For this dataset we chose the \emph{Adadelta} optimizer with learning rate set to 0.9. The dropout-rate was chosen to be 0.5.
When examining the effect of the size of the training data, we can again see  an increasing trend both with respect to accuracy, and (somewhat less pronounced) with respect to F1-score. Further increase in the training dataset would therefore provide slight improvements in the final performance of the model also on this dataset. We used the 300-dimensional version of the publicly available\footnote{\url{https://code.google.com/archive/p/word2vec/}} Google \emph{word2vec} vectors \cite{mikolov2013distributed} to initialize the word embeddings. These vectors are trained on 100 billion words from Google News using the continuous bag-of-words architecture.

\section{Results and Discussion}
\label{results}
In this section we present the results from using both datasets that we presented in Section~\ref{sec:eval-datasets}.

\subsection{Results and Discussion on Twitter Corpus}
\label{sec:results-tweets}

We train our model using the training dataset with development set for tuning the parameters, and then we test the system with the provided testing dataset of 838 test entries, ignoring the ``debatable'' entries.
These results are provided in Table~\ref{table3}.

Recall that there are mainly two components in the proposed approach: (\emph{i}) Sentence modelling using \emph{CNN} and \emph{LSTM}, and (\emph{ii}) Pair-wise word similarity matching. 
Our intuition for using the two models is that both coarse-grained sentence-level and fine-grained word-level information should be important for the paraphrase detection task. In our experiments, we firstly use only sentence modelling architecture to develop the paraphrase detection model. 
We call this experiment the \emph{SentMod Architecture} for paraphrase detection. 
It can be seen from the results in Table~\ref{table3} that the \emph{SentMod} Architecture performs quite well, giving an \emph{F1-score} of 0.692. 
Next, we use the pair-wise word similarity matching model to extract the word-level similarity information-based features. When we use only these features to train the paraphrase model, the model provides an F1-score of 0.702. We call these features the \emph{pair-wise} features. 

Further, we augment these word-level pair-wise features with the sentence-level features extracted using the \emph{SentMod Architecture}, and feed it to train the proposed deep learning model for paraphrase detection task. We call the architecture for this model \emph{DeepParaphrase Architecture}. The experimental results show the significant improvement in the performance of the paraphrase detection task. Specifically, it gives an \emph{F1-score} of 0.742 (an improvement of 7.2 percentage points). 
It also shows that the pair-wise word similarity information in fusion with sentence-level similarity information provides good performance for paraphrase detection task. 
Finally, we augment two additional features, namely the overlap features and similarity features (items \ref{item:similarity-feature} and \ref{item:overlap-feature} in the description in Section~\ref{statsF}). 
This gives an additional improvement in the performance of the model, resulting in an F1-score of 0.751, which is significantly better than other existing methods for paraphrase detection on the Twitter dataset. We refer to this final model as the \emph{AugDeepParaphrase} model.

\begin{table}[htbp]
\caption{\label{font-table2} Results on SemEval 2015 Twitter dataset. }
\label{table3}
\begin{center}
\begin{tabular}{ |l| p{1.6cm} | p{1.6cm} | p{1.6cm} | }
\hline \bf Model  & \bf Precision & \bf Recall  & \bf F1-score \\ 

\hline
SentMod Architecture  & 0.725 & 0.663 & 0.692 \\
\hline
Pair-wise Features   & 0.708 & 0.697 & 0.702 \\
\hline
DeepParaphrase Architecture & 0.753 & 0.731 & 0.742 \\
\hline
AugDeepParaphrase & 0.760 & 0.742 & 0.751 \\

\hline
 
\hline
\end{tabular}
\end{center}

\end{table}

The comparison between the proposed method and existing state-of-the-art methods is provided in Table~\ref{table4}. Firstly, we compare the results of the proposed approach with the best methods on clean text Microsoft Paraphrase dataset, and then with the state-of-the-art methods on noisy Twitter dataset.  \citet{Guo:2012} proposed a weighted textual matrix factorization method for paraphrase detection based on modeling the semantic space of the words that are  present or absent in the sentences. 
Their model uses WordNet, OntoNotes, Wiktionary, and the Brown corpus.
Their approach performed quite well on MSRP dataset, but provide worse results on Twitter dataset. \citet{Das:2009:PIP} used logistic regression based classifier based on simple n-gram features and overlapping features which shows competitive results on MSRP dataset. \citet{DBLP:conf/emnlp/JiE13} presented a state-of-the-art model for paraphrase detection on MSRP dataset which is still the best known performance on clean text. However, it can be seen from the results presented in Table~\ref{table4} that their method performed  worse than other methods on the Twitter data.

\begin{table}[htbp]
\caption{\label{font-table3} Comparison with state-of-the-art of results on SemEval 2015 Twitter dataset. }
\label{table4}
\begin{center}
\begin{tabular}{ |l| p{1.6cm} | p{1.6cm} | p{1.6cm} | }
\hline \bf Model  & \bf Precision & \bf Recall  & \bf F1-score \\
\hline
Random & 0.208 & 0.500 & 0.294 \\
\hline
\citet{Guo:2012}  & 0.583  & 0.525  & 0.655\\
\hline
\citet{Das:2009:PIP} & 0.629 & 0.632  & 0.630 \\ 
\hline
\citet{DBLP:conf/emnlp/JiE13} & 0.664 & 0.628 & 0.645 \\
\hline
\citet{Eyecioglu}  & 0.680  & 0.669 & 0.674 \\
\hline
\citet{Zarrella2015} &  0.569 & 0.806 & 0.667 \\
\hline
\citet{Zhao2015} & 0.767 & 0.583 & 0.662 \\
\hline
\citet{Vo}   & 0.685 & 0.634 & 0.659 \\
\hline
\citet{Karan2015} & 0.645 & 0.674 & 0.659\\
\hline
\citet{Xu_2014}  &  0.722 & 0.726   &  0.724  \\
\hline
\citet{Dey2016}  &  0.756 & 0.726   &  0.741  \\
\hline
\bf AugDeepParaphrase  &   0.760 & 0.742   & \bf  0.751 \\
 
\hline
\end{tabular}
\end{center}

\end{table}

Considering the SemEval 2015 Twitter dataset, Table~\ref{table4} also shows the comparison of our approach against the state-of-the-art methods. As can be observed, the results from this comparison, our approach outperforms all related methods with respect to the F1-score. 
The main reason for this is that our approach leverages the semantic information at both coarse-grained sentence-level features and fine-grained word-level features for detecting paraphrases on tweets. 
The ensemble-based method proposed by \citet{Zarrella2015} obtained higher recall as compared to our results, but our model gave higher overall F1-score. While the method suggested by \citet{Zhao2015} got slightly higher precision as compared to proposed approach, our approach is superior wrt.\ F1-score.

In conclusion, the state-of-the-art algorithms, that perform well when trained on clean texts, do not necessarily work very well for noisy short texts, and vice versa. In contrast to this, our approach is robust in the sense that it performs well on both types of datasets. More specifically, it outperformed all the existing methods when applied on noisy texts, and produced very competitive results against the state-of-the-art methods on clean texts.


Next, we analyze the misclassifications on test data using the proposed approach. Some example tweets pairs including both correct and incorrect detection by our model are reported in Table~\ref{tab:examples-tweet-pairs}. We show some examples from the test data which cases our method could correctly classify. For example, our proposed approach could correctly identify the tweet pair, \emph{``Terrible things happening in Turkey''} and \emph{``Children are dying in Turkey''} as ``paraphrase''. It could understand the semantic meaning, despite the fact that the pair only has one common word. Similarly, the proposed approach could determine correct label as ``non-paraphrase'' for the sentence pairs on row 2 and 3 in Table~\ref{tab:examples-tweet-pairs}, although the sentence-pairs have several words in common. 

Nevertheless, there are several examples where it has been difficult to provide correct classifications. Consider, for example, the tweet pair no.~4 in Table~\ref{tab:examples-tweet-pairs}. 
Our approach determines this pair as         ``non-paraphrase'', which is incorrect according to the gold-standard annotation. The two tweets do not share many words, and common-sense knowledge is required to understand that a person who has won lots of trophies and prizes should be respected rather than hated. 
Another similar example is the tweet pair no.~5 
The gold-standard annotation for this pair is that it is a paraphrase. 
To correctly classify this pair, the system needs to know  that if a person is genius, then it is obvious that he/she would be able to write well.  

\begin{table}[!ht]
\caption{Examples of tweet pairs from the Twitter Paraphrase Corpus.}
\label{tab:examples-tweet-pairs}
\small
\begin{center}
\begin{tabular}{p{1cm} L{3.5cm} L{3.5cm} p{1.6 cm} p{1.6cm} p{1.2cm}}
\toprule  
\bf S. No.& \bf Tweet 1  & \bf Tweet 2 & \bf Gold\newline annotation & \bf Prediction  & \bf Remark \\ 
\midrule
1 &Terrible things happening in Turkey  & Children are dying in Turkey  & paraphrase  &  paraphrase  &  correct \\
 \midrule
2 & Anyone trying to see After Earth sometime soon   & Me and my son went to see After Earth last night   & non-paraphrase & non-paraphrase  &  correct \\
 \midrule
3 & hahaha that sounds like me  & That sounds totally reasonable to me   & non-paraphrase & non-paraphrase  &  correct \\
 \midrule

4 & I dont understand the hatred for Rafa Benitez   & Top 4 and a trophy and still they dont give any respect for Benitez  & paraphrase & non-paraphrase & incorrect  \\
 \midrule
5 & Shonda is a freaking genius &  Dang Shonda knows she can write  & paraphrase & non-paraphrase &   incorrect \\
  \midrule
6 & Terrible things happening in Turkey & Be with us to stop the violence in Turkey
 & paraphrase & non-paraphrase &   incorrect \\
 \midrule
7 & I must confess I love Star Wars  & Somebody watch Star Wars with me please  & paraphrase & non-paraphrase &  incorrect \\
  \midrule
8 & Family guy is really a reality show & Family guy is such a funny show  & non-paraphrase & paraphrase &  incorrect \\
 \midrule
9 & I see everybody watching family guy tonight  & I havent watched Family Guy in forever  & non-paraphrase & paraphrase & incorrect \\
 \bottomrule
\end{tabular}
\end{center}

\end{table} 

Finally, consider the pair \emph{``Family guy is really a reality show''} and \emph{``Family guy is such a funny show''}. Our approach identifies this pair as ``paraphrase'', which is wrong according to the gold-standard annotation. The possible reason for this error is the misleading lexical overlap information between the sentences in the pair, that are overshadowed by the few different words. 

To summarize, after looking at the misclassified examples in Table~\ref{tab:examples-tweet-pairs}, there are several cases that could cause our system to fail to correctly classify pairs of tweets. This includes cases where common-sense knowledge is required. What could be learned from the examples is, however, that our proposed approach is able to capture the semantic information from short, noisy texts, which can, in turn, help in correctly classifying pairs that would otherwise be difficult by only looking at the syntactic contents.

\subsection{Results and discussions on MSRP dataset }
\label{sec:results-MSRP}

The results of our experiments with the Microsoft Paraphrase dataset are summarized in Table~\ref{tab:results-MRSP}. Firstly, we extract the coarse-grained sentence-level features with the \emph{SentMod Architecture} and further feed to train paraphrase detection model. As can be observed in Table~\ref{tab:results-MRSP}, this architecture gives an accuracy of 74.5\% and F1-score of 81.5\%. Next, we evaluate the \emph{pair-wise} features to train the paraphrase detection model, these features individually provide 81.9\% F1-score. Further, we fuse these pair-wise features with the sentence-level features extracted using the SentMod Architecture to train the paraphrase detection model. This \emph{DeepParaphrase Architecture} provides a significant improvement in the performance. With this deep learning model we obtain the accuracy of 77.0\% and F1-score of 84.0\%. The final paraphrase model, \emph{AugDeepParaphrase} model, is built by including the additional features described in Section~\ref{statsF}. Here, we see an improvement of  0.5\% in F1-score. Overall, the experimental results show that both sentence-level semantic information and word-level similarity information are important for paraphrase detection task.

\begin{table}[htbp]
\caption{Results on MSRP dataset. }
\label{tab:results-MRSP}

\begin{center}
\begin{tabular}{ |l | p{1.5cm} |  p{0.9cm} | }
\hline \bf Model  & \bf Accuracy & \bf F1 \\
\hline
SentMod Architecture & 74.5 & 81.5 \\
\hline
Pair-wise Features & 74.8 & 81.9 \\
\hline
DeepParaphrase Architecture & 77.0 & 84.0  \\
\hline
AugDeepParaphrase & 77.7  & 84.5 \\

\hline
\end{tabular}
\end{center}

\end{table}

As with the Twitter dataset, we also compared our approach with several related methods. We present the results of the experiments in Table~\ref{table8}, in which we report the measured accuracy and the F1-scores. The experimental results show that the proposed approach outperforms all the related methods, except quite recent method by \citet{Wang2016SentenceSL}. As discussed in Section~\ref{related}, they also employ a neural network-based approach.
Nevertheless, the large number of options introduced in the final model, such as: the
semantic matching functions \emph{\{max, global, local-l}\}, decomposition operations \emph{\{rigid, linear, orthogonal\}} and filter types \emph{\{unigrams, bigrams, trigrams\}}, makes it less applicable to re-implement or scale for other datasets or other similar problems. In contrast, we have developed our approach to be more robust and generic, such that it can easily be applied for other datasets.

\begin{table}[!ht]
\caption{Experimental results for paraphrase detection on MSRP corpus. }
\label{table8}
\begin{center}
\begin{tabular}{|l|l|l|}
\hline
\bf Model  & \bf Accuracy  & \bf F1 \\
\hline
All positive (Baseline) & 66.5 & 79.9 \\
\hline
\citet{Socher:2011} & 76.8 & 83.6 \\
\hline
\citet{DBLP:conf/emnlp/JiE13}  & 77.8  & 84.3 \\ 
(Inductive setup)  & 	 &   \\ 
\hline
\citet{NIPS2014_5550} ARC- I & 69.6 & 80.3 \\ 
\hline
\citet{NIPS2014_5550} ARC-II & 69.9 & 80.9 \\ 
\hline
\citet{Madnani:2012:RMT} & 77.4 & 84.1 \\ 
\hline
\citet{Eyecioglu}  & 74.4 & 82.8\\
\hline
\citet{El-Alfy:2015} & 73.9 & 81.2\\
\hline
\citet{Wang2016SentenceSL} & \bf 78.4  & \bf 84.7 \\
\hline
\citet{Dey2016} & - & 82.5 \\
\hline
\citet{FERREIRA201859}   & 74.08 & 83.1 \\
\hline
\bf AugDeepParaphrase model  & 77.7   &  84.5 \\
\hline
\end{tabular}
\end{center}

\end{table}



In \cite{DBLP:conf/emnlp/JiE13}, the authors reported the best results as 80.4\% accuracy and 85.9\% F1-score on this dataset. However, to achieve these results, they seemed to have relied on using testing data with training dataset to build the model. They called it a form of transductive learning, in which they assumed that they have access to a test set. In contrast, in our approach, the test data is kept totally disjoint from the training process. Using the same experimental setup -- i.e., applying ``Inductive'' setup without using test data in training the model -- the approach by \citet{DBLP:conf/emnlp/JiE13} gives an accuracy of 77.8\% and F1-score of 84.3\%, which is very close  to the results of our approach. 

Focusing on the performance of our approach in relation to the existing methods, our experimental results
show that our approach produces competitive results, achieving accuracy of 77.7\%  and $F1$ score of 84.5\%. More importantly, we achieved these with less extra annotated resources and no special training strategy, compared to the current state-of-the-art methods.


Table~\ref{tab:example-sentences-MSRP} shows some examples of sentence pairs that our approach has classified both correctly and incorrectly. Sentence pair no. 1 
was correctly classified as ``paraphrase'', even though the sentences do not have many words in common.
Sentence pair no. 2  
was correctly classified as ``non-paraphrase'', even though the two sentences have four words in common words and share the context. 
Conversely, sentence pair no. 5 
was incorrectly predicted as ``paraphrase''. This pair is difficult to classify correctly for humans. 
Sentence pair no. 6 
was incorrectly classified as ``non-paraphrase''. The main reason for this misclassification is the presence of possibly rare words, such as incredulous, jeopardize, endanger, which seemed to have made this sentence pair hard to classify. 

In summary, it seems that our proposed approach is able to capture the semantic information from clean texts, just as it was when analyzing tweets. 
This can, in turn, help in correctly classifying pairs that would otherwise be difficult by only looking at the syntactic contents. There are, however, cases that are hard to classify due to both the lack a complete vocabulary and common-sense knowledge.

\begin{table}[!ht]
\caption{Example sentence pairs from MSRP Paraphrase Corpus.}
\label{tab:example-sentences-MSRP}
\small
\begin{center}
\begin{tabular}{ l L{2.8cm} L{2.8cm} p{1.7 cm} p{1.7cm} p{1.2cm}}
\toprule  
\bf S. No.& \bf Sentence 1  & \bf Sentence 2 & \bf Gold\newline annotation & \bf Prediction  & \bf Remark \\
  
   \midrule
 1 & 	Ricky Clemons' brief, troubled Missouri basketball career is over.   & Missouri kicked Ricky Clemons off its team, ending his troubled career there.   & paraphrase & paraphrase  &  correct \\
\midrule
2 & But 13 people have been killed since 1900 and hundreds injured.  & Runners are often injured by bulls and 13 have been killed since 1900.  & non-paraphrase  &  non-paraphrase  &  correct \\

 \midrule
3 &  I would rather be talking about positive numbers than negative.    & But I would rather be talking about high standards rather than low standards.  & paraphrase & paraphrase  &  correct \\

 \midrule
4 & The tech-heavy Nasdaq composite index shot up 5.7 percent for the week.   & The Nasdaq composite index advanced 20.59, or 1.3 percent, to 1,616.50, after gaining 5.7 percent last week.  & non-paraphrase & paraphrase & incorrect  \\
 \midrule
 
5 & The respected medical journal Lancet has called for a complete ban on tobacco in the United Kingdom.  &  A leading U.K. medical journal called Friday for a complete ban on tobacco  prompting outrage from smokers groups. & non-paraphrase & paraphrase &   incorrect \\
  \midrule
  
6 & Mrs. Clinton said she was incredulous that he would endanger their marriage and family. &	She hadn't believed he would jeopardize their marriage and family. & paraphrase & non-paraphrase &   incorrect \\


\bottomrule
\end{tabular}
\end{center}
\end{table}

\section{Conclusions}
\label{conclusions}
 
In this paper, we introduced a robust and generic paraphrase detection model based on deep neural network model, which performs well on both user-generated noisy short texts such as tweets, and high-quality clean texts.
We proposed a pair-wise word similarity model, which can capture fine-grained semantic corresponding information between each pair of words in given sentences. In addition, we used a hybrid deep neural network that extracts coarse-grained information by developing best semantic representation of the given sentences based on CNN and LSTM.  
The model that we developed consisted of both sentence modelling and pair-wise word similarity matching model. As discussed in this paper, this model proved to be useful for  paraphrase detection. In our evaluation, we included a comprehensive comparison against state-of-the-art approaches. This showed that our approach produced better results than all the existing approaches, in terms of F1-score, when applied on noisy short-text Twitter Paraphrase corpus; and provided very competitive results when applied on clean texts from the Microsoft Paraphrase corpus. Overall, our experimental results have shown the robustness and effectiveness of the proposed method for paraphrase detection. For future work, we plan to investigate how our method works on related tasks such as question answering, sentence matching and information retrieval. We will also further study how to include more close to common-sense knowledge in our model training.

\section*{References}
\bibliography{references}
\bibliographystyle{elsarticle-num-names}


\end{document}